\documentclass[sigchi-a]{acmart}
\usepackage{booktabs} 
\usepackage{ccicons}  

\usepackage{xspace}
\usepackage{xcolor}
\usepackage{amsmath}

\copyrightyear{2019} 
\acmYear{2019} 
\setcopyright{rightsretained} 
\acmConference[CHI'19 Extended Abstracts]{CHI Conference on Human Factors in Computing Systems Extended Abstracts}{May 4--9, 2019}{Glasgow, Scotland UK}
\acmBooktitle{CHI Conference on Human Factors in Computing Systems Extended Abstracts (CHI'19 Extended Abstracts), May 4--9, 2019, Glasgow, Scotland UK}
\acmDOI{10.1145/3290607.3312965}
\acmISBN{978-1-4503-5971-9/19/05}
 \usepackage{booktabs}

\begin{document}
\title{On How Users Edit Computer-Generated Visual Stories}

\author{Ting-Yao Hsu}
\affiliation{%
  \institution{Pennsylvania State University}
  \city{State College}
  \state{PA}
  \postcode{16801}
  \country{USA} }
\email{txh357@psu.edu}

\author{Yen-Chia Hsu}
\affiliation{%
  \institution{Carnegie Mellon University}
  \city{Pittsburgh}
  \postcode{PA}
  \country{USA}}
\email{yenchiah@andrew.cmu.edu}

\author{Ting-Hao (Kenneth) Huang}
\affiliation{%
  \institution{Pennsylvania State University}
  \city{State College}
  \postcode{PA}
  \country{USA}}
\email{txh710@psu.edu}

\renewcommand{\shortauthors}{F. Author et al.}

\newcommand{\kenneth}[1]{{\small\color{blue}{\bf\xspace#1 -Kenneth}}}
\newcommand{\hsu}[1]{{\small\color{red}{\bf\xspace#1 -Hsu}}}

\begin{abstract}
A significant body of research in Artificial Intelligence (AI) has focused on generating \textit{stories} automatically, either based on prior story plots or input images.
However, literature has little to say about how users would receive and use these stories.
Given the quality of stories generated by modern AI algorithms, users will nearly inevitably have to \textit{edit} these stories before putting them to real use.
In this paper, we present the first analysis of how human users \textit{edit} machine-generated stories.
We obtained 962 short stories generated by one of the state-of-the-art \textit{visual storytelling} models. For each story, we recruited five crowd workers from Amazon Mechanical Turk to edit it.
Our analysis of these edits shows that, on average, users 
{\em (i)} slightly shortened machine-generated stories, 
{\em (ii)} increased lexical diversity in these stories, and
{\em (iii)} often replaced nouns and their determiners/articles with pronouns.
Our study provides a better understanding on how users receive and edit machine-generated stories, informing future researchers to create more usable and helpful story generation systems.

\end{abstract}


\begin{margintable}
\vspace{13cm}
\end{margintable}

\maketitle

\section{Introduction}
\begin{marginfigure}
    \begin{minipage}{\marginparwidth}
    \centering
    \includegraphics[width=1\marginparwidth]{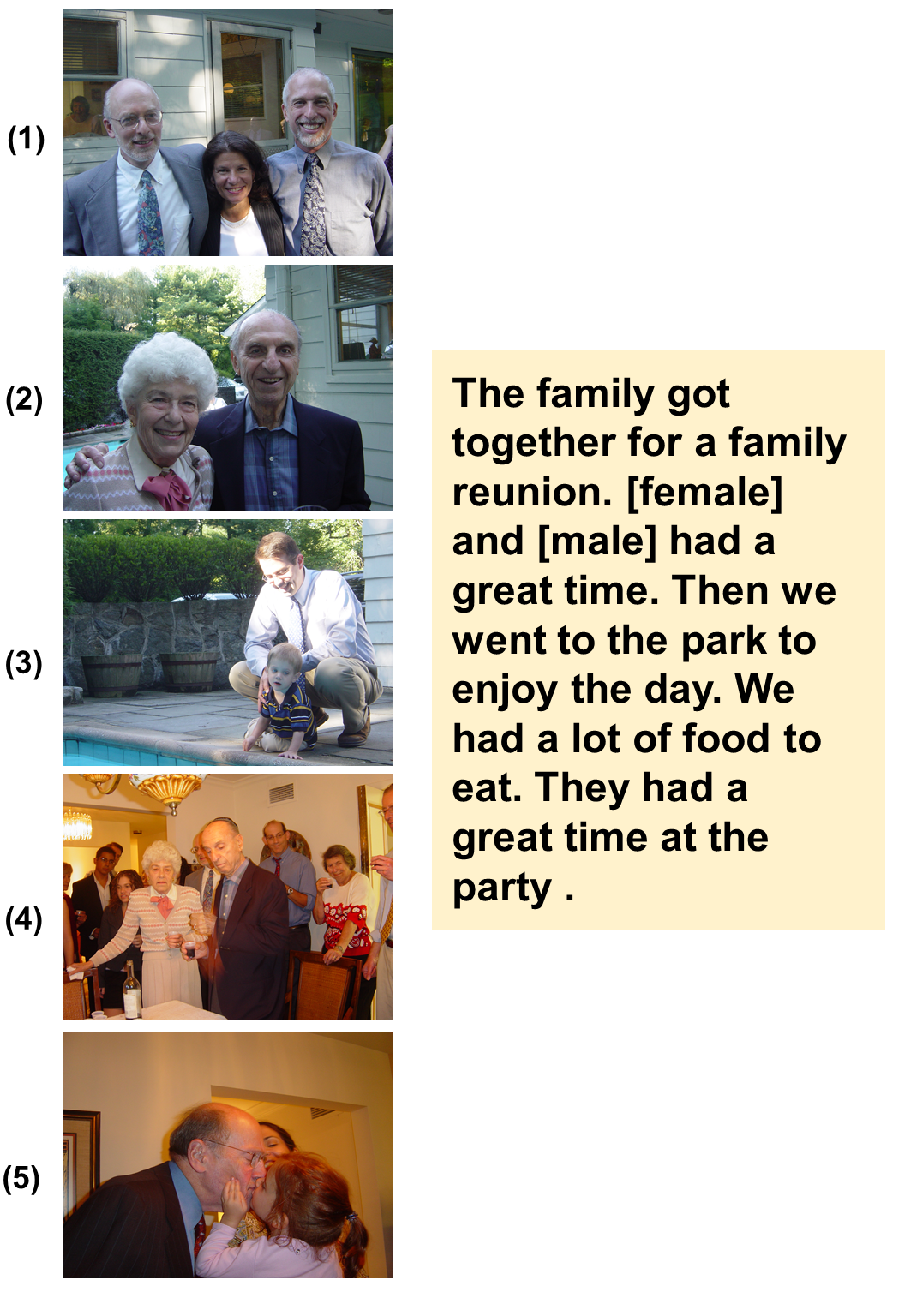}
    \caption{A human-written visual story in the VIST dataset~\cite{huang2016visual}. Each story contains a sequence of five Flickr photos, and a short story written by human workers. (Note that the VIST dataset normalized the story text, {\em e.g.,} replaced all female names, such as ``Amy'' or ``Sarah'', with ``[female]''.)}
    \label{fig:vist-example}
    \end{minipage}
\end{marginfigure}

A large body of research in the AI community has been devoted to generating stories automatically. 
Many prior works aimed to generate stories based on text inputs, such as preceding sentences or writing prompts.
For example, Roemmele {\em et. al} used a recurrent-neural-network architecture to generate stories in a sequence-to-sequence manner~\cite{roemmele2016writing}.
Formulating a story as a sequence of events, Martin {\em et. al} proposed an event representation for neural-network-based story generation~\cite{laraAAAI2018}.
Fan {\em et al.} created a hierarchical model that automatically generates stories conditioning on the writing prompts~\cite{fbStory2018}.
Some other prior work explored generating stories based on images.
For instance, Huang {\em et. al} introduced the \textit{visual storytelling} task, in which the automatic model takes a sequence of photos as input, and generates a short story that narrates this photo sequence~\cite{huang2016visual}.
An example visual story is shown in Figure~\ref{fig:vist-example}.
\textit{StyleNet}~\cite{gan2017styleNet} and \textit{SemStyle}~\cite{mathews2018semstyle} stylized descriptive image captions and made them more ``attractive'', {\em e.g.}, more romantic or humorous.

However, literature has little to say about how users would receive these machine-generated stories and put them to real use.
More specifically, given the quality of stories created by modern automatic models, users will likely need to \textit{edit} these stories for practical uses, for example, sharing them on social media.
In this paper, we focus on the stories generated by modern \textit{visual storytelling} models~\cite{huang2016visual,acl2018wang} and study how people would edit them.
This study allows us to understand how text generation technologies can (or can not) help creative writing.
Some prior work has explored \textit{interactive} computer-supported story writing, in which the system populates suggestions or inspirations for the writer in near real-time when he/she writes the story.
For example, the \textit{Creative Help} system generates suggestions for the next sentence in the process of story writing~\cite{swanson2012say,roemmele2015creative,roemmele2016writing}.
Clark {\em et al.} studied machine-in-the-loop short story writing and concluded that machine intervention should balance between generating coherent and surprising suggestions~\cite{clark2018creative}.
Our study provides a detailed understanding of how users receive and edit machine-generated text in the context of story writing.

\section{Data Preparation}

\paragraph{Machine-Generated Visual Stories}
In the \textit{visual storytelling} task, as introduced by Huang {\em et al.}~\cite{huang2016visual}, the computational model takes a sequence of five photos as input, and then automatically generates a short story describing the photo sequence.
Huang {\em et al.} also released the VIST dataset\footnote{Visual Storytelling Dataset (VIST):\\http://visionandlanguage.net/VIST/}, which contains 81,743 unique Flickr photos in 20,211 sequences, aligned to human-written stories (as shown in Figure~\ref{fig:vist-example}).
Many researchers have proposed approaches to generate visual stories using this dataset.
In this paper, we ran the visual storytelling model released by Wang {\em et al.}~\cite{acl2018wang} on the test set of VIST to obtain machine-generated visual stories.
Wang's approach was one of the state-of-the-art methods that participated in the first Visual Storytelling Challenge~\cite{mitchell2018proceedings} and also the earliest implementation that was made available online.
Among the data in the test set of VIST, we removed the photo sequences containing any photos that have been deleted from Flickr by their owners and only used no more than one photo sequence per Flickr photo album to reduce redundancy.
Eventually, we obtained 962 machine-generated visual stories, and each has a corresponding photo sequence.

\begin{figure}[t]
	\centering
	\includegraphics[width=1\columnwidth]{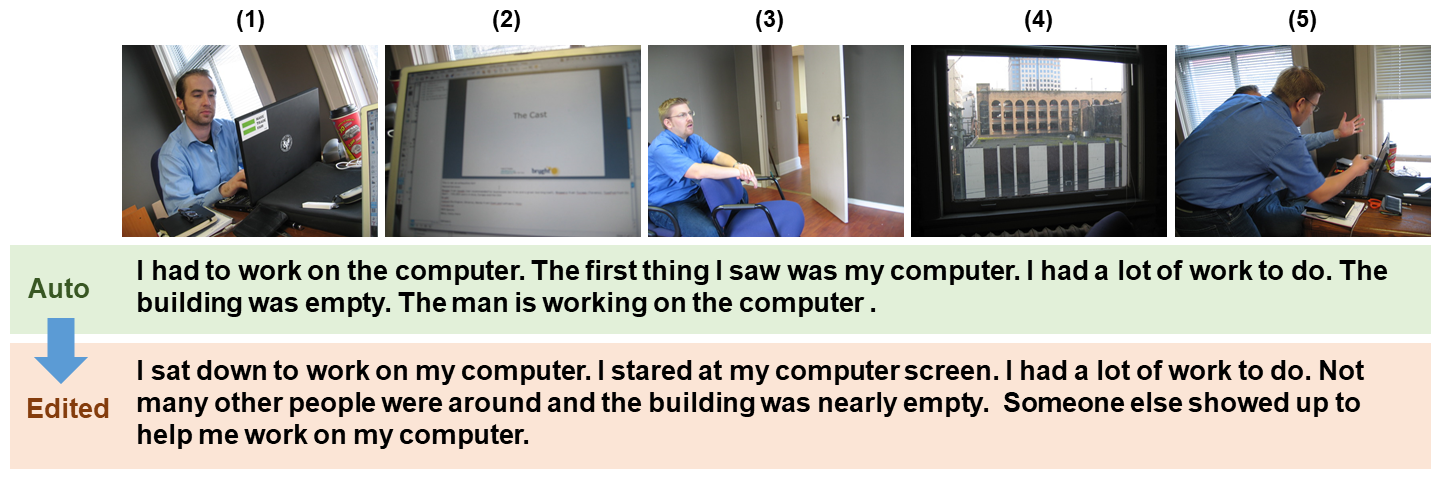}
	\vspace{-2pc}
	\caption{An example of machine-generated visual story (Auto) and its edited version (Edited). This story is generated by the visual storytelling model released by Wang {\em et al.}~\cite{acl2018wang} using the VIST dataset.}
	\label{fig:post-edit-example}
\end{figure}

\begin{marginfigure}
    \begin{minipage}{\marginparwidth}
    \centering
    \includegraphics[width=1.0\marginparwidth]{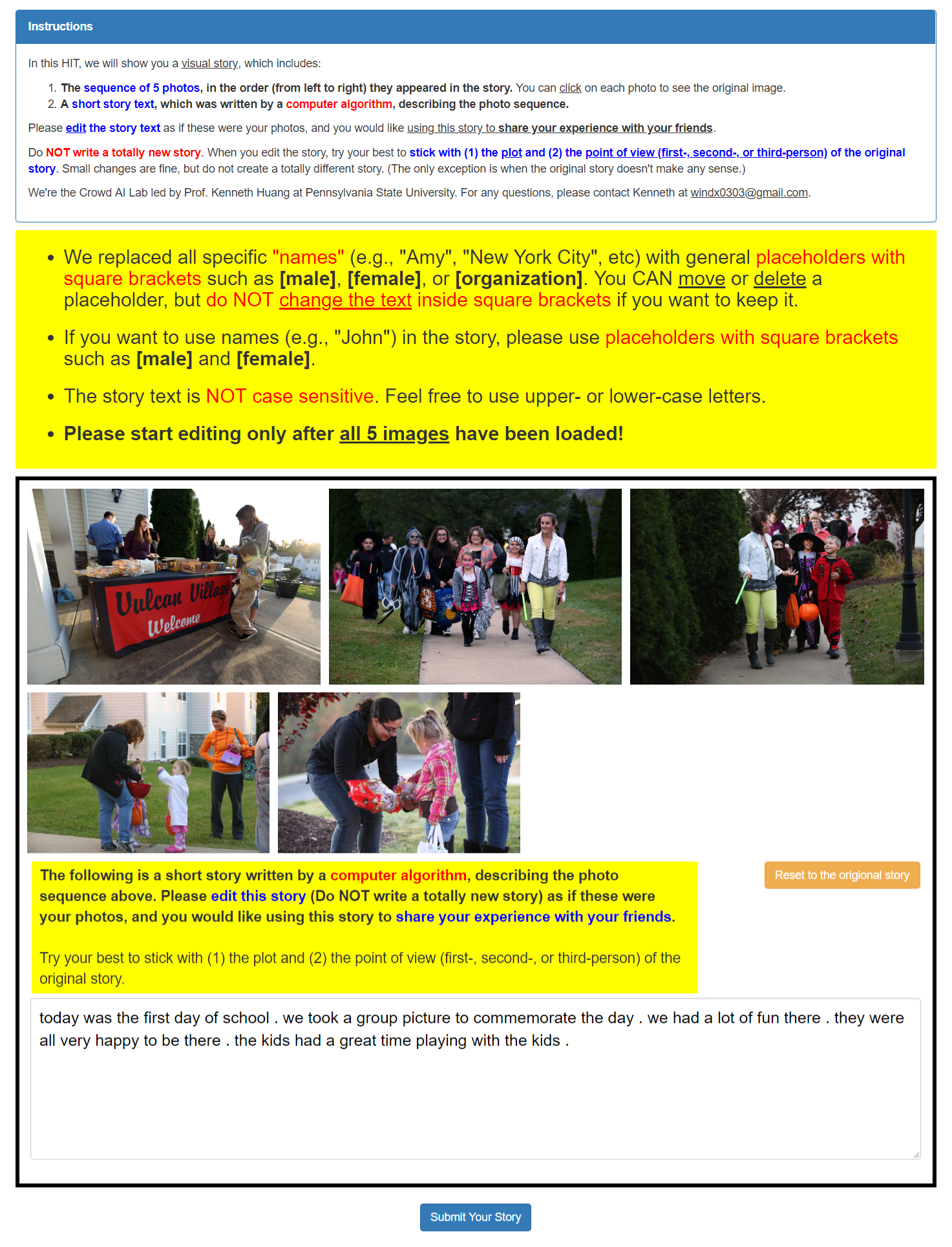}
    \caption{The worker interface for visual story post-editing. The following instruction was used: ``Please edit the story text as if these were your photos, and you would like using this story to share your experience with your friends.'' We also instructed workers to stick with the plot and the point of view (first-, second-, or third-person) of the original story.}
    \label{fig:worker-ui}
    \end{minipage}
\end{marginfigure}

\paragraph{Visual Story Post-Editing}
For each story, we recruited five crowd workers from Amazon Mechanical Turk to edit it, respectively.\footnote{We specified the following qualifications workers must meet to work on our tasks: HIT (Human Intelligence Task) Approval Rate >= 98\%, Number of HITs Approved >= 3000, and location = US. 
The Adult Content Qualification was also used.}
The following instruction was used to guide workers: ``Please edit the story text as if these were your photos, and you would like using this story to share your experience with your friends.''
We instructed workers to stick with the plot and the point of view (first-, second-, or third-person) of the original story so that workers will not abandon the machine-generated story and write a new one from scratch.
The goal of this study is to understand how people edit machine-generated stories. Further research is required to explore factors, such as how coherent the story is, or which level of details does the story provide, that could affect user' willingness to edit a story or abandon it.
The worker interface is shown in Figure~\ref{fig:worker-ui}.
The photo sequence was also displayed on the interface.
The price of each task was \$0.12.
As a result, 197 workers generated 962 $\times$ 5 = 4,810 edited stories.
Figure~\ref{fig:post-edit-example} shows an example of a generated visual story, before and after human editing.

\section{Analysis}


We analyzed the length, type-token ratio (TTR), and the part of speech tags of pre- and post-edited visual stories.
We used the NLTK toolkit to segment sentences, tokenize sentences into words (known as ``tokens''), and tag part-of-speech (POS) for each word~\cite{bird2009natural}.
It is noteworthy that the VIST dataset normalized the story text, {\em e.g.,} replaced all female names such as ``Amy'' and ``Sarah'' with ``[female]''.
We replaced all ``[female]'' with ``Amy'' and ``[male]'' with ``Tom'' for our analysis.
The detailed results and discussions are as follows.


\subsection{Edited stories are slightly shorter.}
On average, the edited visual stories are shorter significantly (paired t-test, $p<0.001$, $N=4810$).
We calculated the number of tokens (also known as ``sentence length'') of each story.
An automatic story contains an average of 43.02 tokens (punctuation marks included, SD = 4.96), and an edited story contains an average of 41.85 tokens (SD = 9.70).
This comparison indicates that users on average eliminated 1.17 tokens.
Meanwhile, we noticed that the edited stories' lengths resulted in a higher sample standard deviation (SD), which suggested that human-written stories are more diverse regarding story length.

\subsection{Edited stories have a higher lexical diversity.}
\begin{marginfigure}
    \vspace{-20pc}
    \begin{minipage}{\marginparwidth}
    \centering
    \includegraphics[width=1.0\marginparwidth]{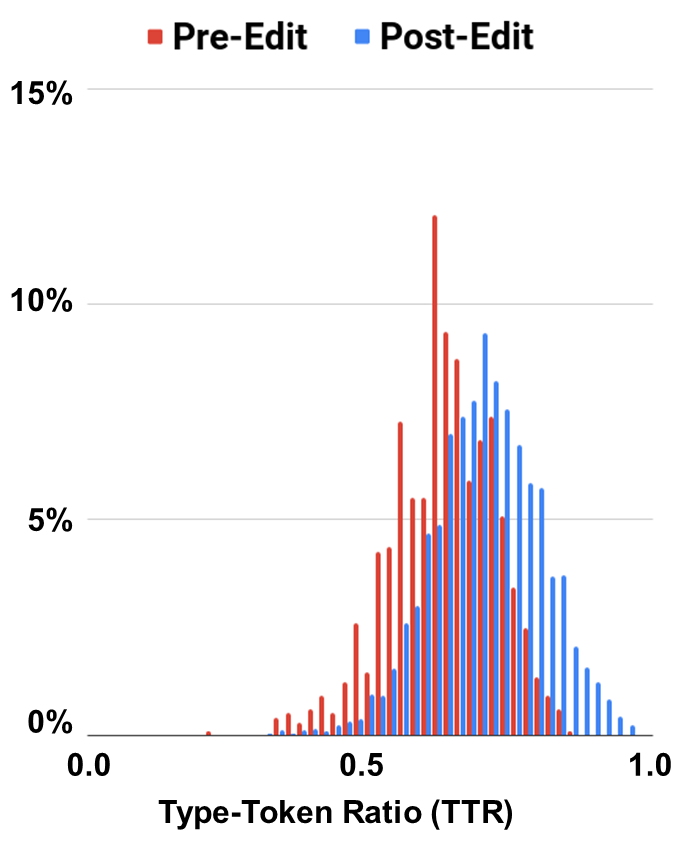}
    \caption{The normalized frequency histogram of TTR of pre- and post-edited stories. A higher TTR indicates a higher lexical diversity. This result suggests that users increased the lexical diversity in machine-generated stories.}
    \label{fig:ttr-histo}
    \end{minipage}
\end{marginfigure}

We also studied the lexical diversity of these storied.
Type-token ratio (TTR) is the one of most commonly used index of lexical diversity~\cite{1af8cfb45ecf4823b5e14c69b80d4d5a}, which is calculated as the ratio between the number of different words (types) $N_{type}$ and the total number of words (tokens) $N_{token}$ of a text unit, {\em i.e.,} $TTR = N_{type}/N_{token}$. 
A higher TTR indicates a higher lexical diversity.
Per our analysis, the average TTR for an automatic story is 0.62 (SD = 0.09), while the average TTR for an edited story is 0.72 (SD = 0.06).
This difference is statistically significant (paired t-test, $p<0.001$, $N=4810$).
This result suggested that users reduced the word redundancy and increased the lexical diversity in machine-generated stories.
The normalized frequency histogram of TTR of pre- and post-edited stories are shown in Figure~\ref{fig:ttr-histo}.

\subsection{Nouns (NOUN) with determiners/articles (DET) are often replaced by pronouns (PRON).}
We also analyzed the parts of speech of the words in each story.
The universal part-of-speech tagset\footnote{A Universal Part-of-Speech Tagset:\\ https://www.nltk.org/book/ch05.html} provided by NLTK toolkit was used.
The average number of tokens of each POS in a story is shown in Table~\ref{tab:pos-stats}.
We observed that the ``DET'' (determiner, article) tag contributes the most to the reduction of story length, while the number of ``PRON'' (pronoun) tags increased the most.
In order to understand the possible reasons behind this phenomenon, we observed the stories that had fewer DET tags but more PRON tags after post editing.
We found that the nouns with determiners and articles (DET) are often replaced by pronouns (PRON).
The following is a typical example, where the replacement texts are highlighted in red.

\begin{quote}
\textbf{[pre-edit]} \textcolor{red}{the car} was parked in the middle of the road . ...
\end{quote}

\begin{quote}
\textbf{[post-edit]} \textcolor{red}{we} drove home and parked by lots of other cars . ...
\end{quote}

The following is another example:

\begin{quote}
    \textbf{[pre-edit]} the group of friends went to the camp site to see what was going on . we had a lot of food to eat . \textcolor{red}{the kids} had a great time . \textcolor{blue}{we had a lot of fun . we had a lot of fun .}
\end{quote}

\begin{quote}
\textbf{[post-edit]} a group of friends went camping together . they took plenty of food to eat . \textcolor{red}{they} had a great time together . \textcolor{blue}{everyone had a lot of fun .}
\end{quote}

We also noticed that the ``.'' (punctuation marks) tags reduced significantly.
This might be caused by the fact that machine-generated stories can be repetitive or too general, so users often merged two sentences or simply removed one of them.
For instance, in the example above (the text highlighted in blue), the user decided to remove one sentence and thus one period mark (``.'') was eliminated.

\begin{table}[t]
\small
\centering
\caption{Average number of tokens with each POS tag per story. The universal part-of-speech tagset provided by NLTK toolkit was used. $\Delta$ denoted the differences in average token numbers between between post- and pre-edit stories. The ``DET'' (determiner, article) tag contributes the most to the reduction of story length, and the ``.'' (punctuation marks) tag contributes the second.}
\begin{tabular}{@{}llllllllllllr@{}}
\toprule
 & \textbf{ADJ} & \textbf{ADP} & \textbf{ADV} & \textbf{CONJ} & \textbf{DET} & \textbf{NOUN} & \textbf{NUM} & \textbf{PRT} & \textbf{PRON} & \textbf{VERB} & \textbf{.} & \textbf{Total} \\ \midrule
\textbf{Pre-Edit} & 3.09 & 3.46 & 1.92 & 0.50 & 8.11 & 10.00 & 0.02 & 1.62 & 2.14 & 7.00 & 5.16 & 43.02 \\
\textbf{Post-Edit} & 3.13 & 3.39 & 1.85 & 0.85 & 7.15 & 9.81 & 0.05 & 1.58 & 2.31 & 7.05 & 4.69 & 41.85 \\ \midrule
\textbf{$\Delta$} & 0.04 & -0.07 & -0.07 & 0.35 & -0.96 & -0.20 & 0.03 & -0.04 & 0.17 & 0.04 & -0.47 & -1.17 \\ \bottomrule
\end{tabular}
\label{tab:pos-stats}
\end{table}

\section{Discussion \& Future Work}
Our analysis on the machine-generated and post-edited visual stories shows that, on average, users 
{\em (i)} slightly shortened machine-generated stories, 
{\em (ii)} increased lexical diversity, and
{\em (iii)} often replaced nouns and their determiners/articles with pronouns, and merged or eliminated sentences.
The higher-level theme emerged in these observations are \textit{text repetition} in machine-generated stories.
Generating redundant text is a known problem for many neural-network-based visual storytelling models, while some approaches~\cite{hsu2018using} suffer from this problem more than others~\cite{kim2018glac}.
Our analysis indicates that, for the visual stories generated by Wang {\em et al.}~\cite{acl2018wang}, a significant part of human editing effort is to reduce word redundancy and increase lexical diversity.
Furthermore, users frequently replaced nouns with pronouns suggests that the storytelling model should better understand which entities ({\em e.g.,} human, building, animal) have been mentioned and thus can be called using pronouns ({\em e.g.,} she, it).
In the future, we will develop algorithms to learn from these edits to improve existing machine-generated stories.
We will also study what types of machine-generated texts are more helpful in assisting users to compose short stories, aiming at building a human-centered computer-supported storytelling system.

\section{Acknowledgements}
We thank Jiawei Chen for her help.
We also thank the workers on Mechanical Turk who participated in our experiments.

\bibliography{main.bib}
\bibliographystyle{ACM-Reference-Format}

\end{document}